\documentclass[intlimits,twoside,a4paper]{article}

\usepackage{amsmath,amssymb}
\usepackage{graphicx}
\usepackage{wrapfig}

\usepackage[T2A]{fontenc}
\usepackage[cp1251]{inputenc}
\usepackage[eqsecnum]{cmpj2}

\issue{2012}{15}{3}{33702}
\doinumber{10.5488/CMP.15.33702}

\title[Off-central acceptor impurity in a spherical quantum dot]%
{Off-central acceptor impurity in a spherical \\quantum dot}
\author[V.I.~Boichuk \textsl{et al.}]{V.I.~Boichuk, R.Ya.~Leshko\thanks{E-mail: leshkoroman@meta.ua}\,, I.V.~Bilynskyi, L.M.~Turyanska}
\address{Department of Theoretical Physics, Ivan Franko Drohobych
State Pedagogical University, \\ 3 Stryiska Str., 82100 Drohobych, Ukraine
}

\date{Received February 20, 2012}

\authorcopyright{V.I.~Boichuk, R.Ya.~Leshko, I.V.~Bilynskyi, L.M.~Turyanska, 2012}

\begin{document}

\maketitle

\begin{abstract}
The hole energy spectrum with the ion of an acceptor impurity in the quantum dot has been calculated using the spherical approximation of the multiband Luttinger model. The dependence of the hole energy levels on the impurity location in the quantum dot has been studied. The effect of the impurity location on the dipole momentum and the oscillator strength has been analyzed. The hole interlevel absorption coefficient has been calculated.
\pacs 73.21.La, 71.55.-i, 78.20.Ci
\end{abstract}

\section{Introduction}

For the last decade numerous theoretical and experimental works related to quantum dots (QD's) have been carried out~\cite{Klim,Wilf,Tkach1,Bond,Wei,Boich1}. The physical properties of the spherical QD, such as a dipole transition, oscillator strength, and optical absorption coefficient can significantly depend on the presence of impurities in QD's. Today,  a lot of theoretical and experimental works are available, where the effect of impurities on the QD optical parameters has been studied. Most theoretical works analyzed hydrogenic (donor or acceptor) impurities~\cite{Boich1,Polup,Tkach2,Boic2,Boic3}  located in the center of the QD or quantum anti-dot~\cite{Holov}. For a donor impurity within the effective mass approximation, an exact solution of Schr\"odinger equation has been obtained.
As concerns the acceptor impurity, the multiband model of the valency band has been used, such as Luttinger model~\cite{Lutt,Bald}. In the case of a harsh change of the heterosystem dielectric permittivity, the theory has been constructed which makes it possible to regard the effect of polarization charges on hydrogenic impurity~\cite{Boich1,Boic2,Boic3}.

In general, the impurity can be anywhere in the QD or even outside. In this case, the Schr\"odinger equation does not have an exact solution. That is why variational methods are used in calculating the ground and exited states of an impurity. It was found that the impurity shift from the QD center is caused by the splitting of degenerated levels~\cite{Zhu,Boic4}. The number of splitting levels is equal to the number of magnetic number values. The energy spectrum of an impurity electron has been so far studied in the cubic QD using a variational method~\cite{Li}. General properties of an impurity in the spherical QD in the presence of an electric field were analyzed in~\cite{Nasria} using a plane wave basis. The Stark effect was researched. An ellipsoidal QD with off-central impurity in the parabolic potential well is considered in~\cite{Sadeghi}. Impurity eigenvalues were defined by expanding the exact wave function over exact functions of the system with a central impurity. The binding energy was presented as a function of the impurity position and ellipticity constant.

Despite a large number of QD works, we have not come across the works in which the off-central acceptor impurity was analyzed within the Luttinger model for cubic crystals. Given that the valence band of most semiconductors of cubic symmetry is degenerate in the center of the Brillouin zone, the Luttinger model should be employed for an adequate theoretical analysis of hole and acceptor states. Moreover, the problem of the off-central impurity is important because the impurity can be located on the surface of the QD. It is known that in many experiments nanoparticles were obtained formed on the surface nanoclusters~\cite{Bond}. Such problems are an important step for a further construction of the theory of surface impurity defects on the QD.

As noted before, impurities can change optical properties, in particular the absorption coefficient, which is caused by the hole or electron interlevel transitions. The interest to intraband interlevel transitions is caused by the possibility to use those transitions in the construction of terahertz radiation detectors~\cite{Wei}, because the transition energy is in the terahertz range. Particularly, this urges the study of interlevel transitions in the QD with impurities~\cite{Yakar,Xie1,Xie2,Rezaei1,Vahdani,Rezaei2}. In particular, in~\cite{Boic4} the dependence of the absorption coefficient on the impurity position in the QD was studied. It was shown that the displacement of a donor impurity ion from the QD center causes shift of the absorption band into the low-energy range. Although in these works interlevel transitions in the QD with the donor impurity are described quite in detail, but optical transitions caused by the off-central acceptor impurity have not been researched.

Therefore, the work is aimed at
\begin{itemize}
  \item calculating the energy spectrum of the off-central acceptor impurity within Lattinger multiband model;
  \item studying the QD optical parameters (dipole momentum, oscillator strength, absorption coefficient) with the off-central acceptor impurity;
  \item analyzing the obtained results with respective results of the donor impurity.
\end{itemize}

\section{Eigenvalues and eigenfunctions of the off-central acceptor impurity}

\subsection{Formulation of the problem and its solution}

We consider a spherical QD heterostructure with an acceptor impurity which is located on the distance $D$ from the center of the QD. The radius of the QD is $a$. Let the heterosystem be constructed of cubic crystals with the large
band gap $E_{\mathrm{g}}$ and strong spin-orbit interaction $\Delta$. Let those crystals be 4-fold degenerated in the $k=0$. Taking this into account, the spherical Luttinger Hamiltonian is given by \cite{Lutt,Bald,Shkov}
\begin{equation}
\label{Hamiltonian}
    {\bf{H}} = \frac{1}{2}\left( {{\gamma _1} + \frac{5}{2}\gamma } \right){{\bf{p}}^2} - \gamma {\left( {{\vec{\bf p}} \cdot {\vec{\bf J}}} \right)^2} + \Pi \left( {\vec r} \right),
\end{equation}
where $\gamma_1$, $\gamma=1/5(2 \gamma_2+ 3 \gamma_3)$ are Luttinger parameters, $\vec{\bf{p}}$ is the momentum operator, $\vec{\textbf{J}}=\vec{i}\textbf{J}_x+ \vec{j}\textbf{J}_y+ \vec{k} \textbf{J}_z$ is the spin operator $j=3/2$. The potential energy of the system is given by
\begin{equation}
\label{potential}
    \Pi \left( r \right) = W\left( {\vec r,\vec D} \right) + U\left( r \right).
\end{equation}
The interaction between the acceptor ion and the hole has been written in the Coulomb form
\begin{equation}
\label{coulomb}
    W\left( {\vec r,\vec D} \right) =  - \frac{{{\re^2}}}{{4\pi {\varepsilon _0}\varepsilon \big| {\vec r - \vec D} \big|}}\,,
\end{equation}
where $\varepsilon_0$ is an electric constant, $\varepsilon$ is a dielectric permittivity of the QD. In this work we consider a heterosystem with a large band mismatch. That is why the confinement potential has been chosen as the infinitely high spherically symmetric potential well
\begin{equation}
\label{confinement}
U\left( r \right) = \left\{
                     \begin{array}{ll}
                            0, & \hbox{$r < a$,}\\
                            \infty, & \hbox{$r\geqslant a$.}
                    \end{array}
\right.
\end{equation}
The exact solution of the Schr\"odinger equation with Hamiltonian (\ref{Hamiltonian}) does not exist. But if the potential (\ref{coulomb}) is neglected, the exact solution can be obtained \cite{Shkov}
\begin{equation}
\label{generalSolution}
    {\psi _{f,M}}\left( {r,\theta ,\varphi } \right) = \sqrt {2f + 1} \sum\limits_{l = f - j}^{f + j} {{{\left( { - 1} \right)}^{l - j + M}}R_f^l\left( r \right)\sum\limits_{{m_l}}^{} {\sum\limits_{{m_j}}^{} {\left( {
    \begin{array}{*{20}{c}}
        l&j&f\\
        {{m_l}}&{{m_j}}&{ - M}
    \end{array}
    }\right){Y_{lm}}\left( {\theta ,\varphi } \right){\chi _{{m_j}}}} } }\,,
\end{equation}
where ${\hbar ^2}f\left( {f + 1} \right)$, ${\hbar ^2}l\left( {l + 1} \right)$, $\hbar M$, $\hbar m$, $\hbar m_j$ are eigenvalues of operators ${\bf F}^2$, ${\bf L}^2$, ${\bf F}_z$, ${\bf L}_z$, ${\bf J}_z$ respectively, $\chi_{m_j}$ are spin functions, $Y_{l,m}$ are spherical harmonics,
$
\left( 
\begin{smallmatrix}
l&j&f\\
{{m_l}}&{{m_j}}&{ - M}
\end{smallmatrix}
\right)
$
are 3-j symbols. Based on the general function~(\ref{generalSolution}), exact solutions in the spherically symmetric field have been obtained in~\cite{Bald,Sheka,Gelmont,Grigoryan} for three types of states
\begin{equation}
\label{IIItypes}
    \left\{ \begin{array}{ll}
    \hspace{-2mm}\psi _{f,M}^{\mathrm{I}} = R_f^{f - 3/2}\left( r \right)\Phi _{f,M}^{f - 3/2}\left( {\theta ,\varphi } \right) + R_f^{f + 1/2}\left( r \right)\Phi _{f,M}^{f + 1/2}\left( {\theta ,\varphi } \right), & f \geqslant  3/2 \ \left( {l = f - 3/2,\,\,f + 1/2} \right),\\[1,5ex]
    \hspace{-2mm}\psi _{f,M}^{\mathrm{II}} = R_f^{f - 1/2}\left( r \right)\Phi _{f,M}^{f - 1/2}\left( {\theta ,\varphi } \right) + R_f^{f + 3/2}\left( r \right)\Phi _{f,M}^{f + 3/2}\left( {\theta ,\varphi } \right), & f \geqslant 3/2 \ \left( {l = f - 1/2,\,\,f + 3/2} \right),\\[1,5ex]
    \hspace{-2mm}\psi _{f,M}^{\mathrm{III}} = R_{1/2}^l\left( r \right)\Phi _{1/2,M}^l\left( {\theta ,\varphi } \right),
& f = 1/2 \ \left( {l = 1,\,\,\,2} \right),
\end{array} \right.
\end{equation}
where $\Phi _{f,M}^l\left( {\theta ,\varphi } \right)$ are spinors which correspond to the spin $j=3/2$. For convenience,  as a unit of length we use the effective Bohr radius ($a_\mathrm{b}^* = 0.53\,{\varepsilon _1}{\gamma _1}$~\AA), and as a unit of energy we use $\mathrm{Ry}^* = {13.6}/{\left({\varepsilon ^2}{\gamma _1}\right)}$~eV which represents the effective Rydberg energy. In this system of units, radial wave functions have the form:
\begin{equation}
\label{IandIIType}
    \left( {\begin{array}{*{20}{c}}
    {R^i_2}\\
    {R^i_1}
    \end{array}} \right) = {A_1}\left( {\begin{array}{*{20}{c}}
    {\frac{{ - {C_2}}}{{1 + {C_1} - \left( {1 + \mu } \right)}}{j_l}\left[ {\chi \sqrt {\left( {1 - \mu } \right)} \frac{r}{a}} \right]}\\
    {{j_{l + 2}}\left[ {\chi \sqrt {\left( {1 - \mu } \right)} \frac{r}{a}} \right]}
    \end{array}} \right) + {A_2}\left( {\begin{array}{*{20}{c}}
    {\frac{{ - {C_2}}}{{1 + {C_1} - \left( {1 - \mu } \right)}}{j_l}\left[ {\chi \sqrt {\left( {1 + \mu } \right)} \frac{r}{a}} \right]}\\
    {{j_{l + 2}}\left[ {\chi \sqrt {\left( {1 + \mu } \right)} \frac{r}{a}} \right]}
    \end{array}} \right),
\end{equation}
where $i=\mathrm{I}, \mathrm{II}$ (the set of numbers determines the type of states), $R_1^i = \left\{ {R_f^{f + 1/2},\ R_f^{f + 3/2}} \right\}$, \linebreak $R_2^i = \left\{ {R_f^{f - 3/2},\ R_f^{f - 1/2}} \right\}$, respectively and coefficients have been written with the use of 6-j symbols
\begin{eqnarray*}
    {C_1} &=& {C_1}\left( {f,l} \right) = \mu \sqrt 5 {\left( { - 1} \right)^{3/2 + l + f}}\left\{ {
        \begin{array}{*{20}{c}}
            l&l&2\\
            {3/2}&{3/2}&f
        \end{array}
    } \right\}\sqrt {\frac{{2l\left( {2l + 1} \right)\left( {2l + 2} \right)}}{{\left( {2l + 3} \right)\left( {2l - 1} \right)}}}\,,\\[1.5ex]
    {C_2} &=& {C_2}\left( {f,l} \right) = \mu \sqrt {30} {\left( { - 1} \right)^{3/2 + l + f}}\left\{ {
        \begin{array}{*{20}{c}}
            {l + 2}&l&2\\
            {3/2}&{3/2}&f
        \end{array}
    } \right\}\sqrt {\frac{{\left( {l + 1} \right)\left( {l + 2} \right)}}{{2l + 3}}}\,,\\[1.5ex]
&&{\left( {{C_1}} \right)^2} + {\left( {{C_2}} \right)^2} = {\mu ^2}, \qquad {C_2}/\mu  > 0, \qquad \mu  = \frac{{2\gamma }}{{{\gamma _1}}}\,.
\end{eqnarray*}
$j_l$ is  the spherical Bessel function of the first type. The solution for the third type of states can be represented by the one spherical Bessel function of the first type. Based on the boundary conditions for an infinitely high potential well, the dispersion equation has been derived, from which the parameter $\chi _{f,l,n}$ has been obtained. Here, $n$ is the number of the solution of dispersion equation for other fixed quantum numbers. Therefore, the energy of the hole is written in the form
\begin{equation}
\label{Ehole}
    {E_{f,l,n}} = {\left( {\frac{{{\chi _{f,l,n}}}}{a}} \right)^2}\left( {1 - {\mu ^2}} \right).
\end{equation}
Since the energy depends on two more quantum numbers, the function (\ref{IIItypes}) can be defined $\psi _{f,M}^{\mathrm{I}} = \psi _{f,M;n,l}^{\mathrm{I}}$; $\psi _{f,M}^{\mathrm{II}} = \psi _{f,M;n,l}^{\mathrm{II}}$; $\psi _{f,M}^{\mathrm{III}} = \psi _{f,M;n,l}^{\mathrm{II}}$. In the case of the presence of an impurity in the QD, the Ritz variational method was used to determine the energy of the system. To construct the variational function, the wave function of the task without the impurity (\ref{IIItypes})--(\ref{IandIIType}) was used. From boundary conditions for the function (\ref{IandIIType}), $A_2$ was expressed by $A_1$
\begin{displaymath}
    {A_2} =  - {A_1}\frac{{{j_{l + 2}}\left( {\chi \sqrt {\left( {1 - \mu } \right)} } \right)}}{{{j_{l + 2}}\left( {\chi \sqrt {\left( {1 + \mu } \right)} } \right)}}\,.
\end{displaymath}
Then, (\ref{IandIIType}) can be written as:
\begin{eqnarray}\label{I-II}
\left( {\begin{array}{*{20}{c}}
{R^i_2}\\
{R^i_1}
\end{array}} \right) &=& {A_1}\left( {\begin{array}{*{20}{c}}
{\frac{{ - {C_2}}}{{1 + {C_1} - \left( {1 + \mu } \right)}}{j_l}\left[ {\chi \sqrt {\left( {1 - \mu } \right)} \frac{\rho }{a}} \right]}\\
{{j_{l + 2}}\left[ {\chi \sqrt {\left( {1 - \mu } \right)} \frac{\rho }{a}} \right]}
\end{array}} \right)\nonumber\\
 &&{}- {A_1}\frac{{{j_{l + 2}}\left[ {\chi \sqrt {\left( {1 - \mu } \right)} } \right]}}{{{j_{l + 2}}\left[ {\chi \sqrt {\left( {1 + \mu } \right)} } \right]}}\left( {\begin{array}{*{20}{c}}
{\frac{{ - {C_2}}}{{1 + {C_1} - \left( {1 - \mu } \right)}}{j_l}\left[ {\chi \sqrt {\left( {1 + \mu } \right)} \frac{\rho }{a}} \right]}\\
{{j_{l + 2}}\left[ {\chi \sqrt {\left( {1 + \mu } \right)} \frac{\rho }{a}} \right]}
\end{array}} \right) = {A_1}\left( {\begin{array}{*{20}{c}}
{{S_i}}\\
{{G_i}}
\end{array}} \right).\qquad
\end{eqnarray}
Using (\ref{I-II}) the ground state variational function (first type of states) was written in the form
\begin{equation}\label{groundWavefunction}
    {\Psi _{\mathrm{I}}} = {A_{\mathrm{I}}}\left[ {{Q_{\mathrm{I}}}\,{S_{\mathrm{I}}}\left( r \right)\Phi _{f,M}^{f - 3/2}\left( {\theta ,\varphi } \right) + {W_{\mathrm{I}}}\,{G_{\mathrm{I}}}\left( r \right)\Phi _{f,M}^{f + 1/2}\left( {\theta ,\varphi } \right)} \right]{\re^{ - {\alpha _{\mathrm{I}}}\sqrt {{r^2} + {D^2} - 2rD\cos \theta } }},
\end{equation}
where $Q_{\mathrm{I}}$, $W_{\mathrm{I}}$ are linear variational parameters, $\alpha_{\mathrm{I}}$ is the variational parameter, $\theta$ is the angle between the direction on the impurity and the hole. The symmetry of the problem makes it possible to choose the coordinate system in the way that the axis $z$ passes through the QD centre and through the ion of the impurity. $A_{\mathrm{I}}$ can be determined from the normalization condition. For the ground state $f = 3/2$, $M =-3/2,-1/2,1/2,3/2$.

The wave function (\ref{groundWavefunction}) in the Schr\"odinger equation was substituted  with the Hamiltonian~(\ref{Hamiltonian}). The obtained expression was multiplied by the Hermitian conjugate function~(\ref{groundWavefunction}) and the final expression was integrated by angle variables. As a result, the functional $F = F\left( {{Q_{\mathrm{I}}},{W_{\mathrm{I}}},{\alpha _{\mathrm{I}}},M} \right)$ was obtained. The functional was minimized and the ground state energy that depends on $\left| M \right|$ was defined. In the case of the central impurity $D=0$, the energy does not depend on $\left| M \right|$. Also, in the case $D=0$, Ritz variational method can be used to calculate the first excited state (second type of states) with the similar wave function:
\begin{eqnarray*}
    {\Psi _{\mathrm{II}}} = {A_{\mathrm{II}}}\left[ {{Q_{\mathrm{II}}}\,{S_{\mathrm{II}}}\left( r \right)\Phi _{f,M}^{f - 1/2}\left( {\theta ,\varphi } \right) + {W_{\mathrm{II}}}\,{G_{\mathrm{II}}}\left( r \right)\Phi _{f,M}^{f + 3/2}\left( {\theta ,\varphi } \right)} \right]{\re^{ - {\alpha _{\mathrm{II}}}r}},
\end{eqnarray*}
which, due to orthogonality spinors $\Phi _{f,M}^l\left( {\theta ,\varphi } \right)$, is orthogonal to the function~(\ref{groundWavefunction}) when $D=0$. As concerns the off-central impurity, the wave function cannot be presented in the form similar to~(\ref{groundWavefunction}). Since the $\cos(\theta)$ is in the exponent, the functions are not orthogonal. In this case, additional parameters should be included for orthogonalization of the functions. Such actions may complicate the minimization of the functional. In this regard, we expressed the off-central impurity wave function by the linear expansion over the wave function without impurity:
\begin{equation}\label{WaveFuctionExpansion}
    \Psi  = \sum\limits_i {{c_i}\psi _i^0}
\,,
\end{equation}
where $\psi _i^0$ is the wave function of the hole in the QD [$W(\vec{r},\vec{D})=0$] which satisfies the Schr\"odinger equation
\begin{displaymath}
    {{\bf{H}}^0}\psi _i^0 = E_i^0\psi _i^0\,.
\end{displaymath}
The solution of this equation is of the form (\ref{groundWavefunction}). The index $i$ denotes the whole set of quantum numbers that characterize the state of the hole. After substituting (\ref{WaveFuctionExpansion})  with Hamiltonian (\ref{Hamiltonian}) into the Schr\"odinger equation, the linear system of equations was obtained regarding  $c_i$:
\begin{equation}\label{linearSystem}
    \sum\limits_i {\left[ {\left( {E_i^0 - E} \right){\delta _{ji}} + {W_{ji}}} \right]{c_i}}  = 0,
\end{equation}
where $W_{ji}$ is the matrix element of the potential energy (\ref{coulomb}) on functions of the problem without impurity. The energy spectrum of the acceptor impurity and coefficients $c_i$  were obtained by  solving the linear system of equations (\ref{linearSystem}) and the normalization condition $\sum\limits_i {{{\left| {{c_i}} \right|}^2}}  = 1$.

\subsection{Analysis of the spectrum of the acceptor impurity}

\begin{figure}[!h]
\centerline{
\includegraphics[width=0.65\textwidth]{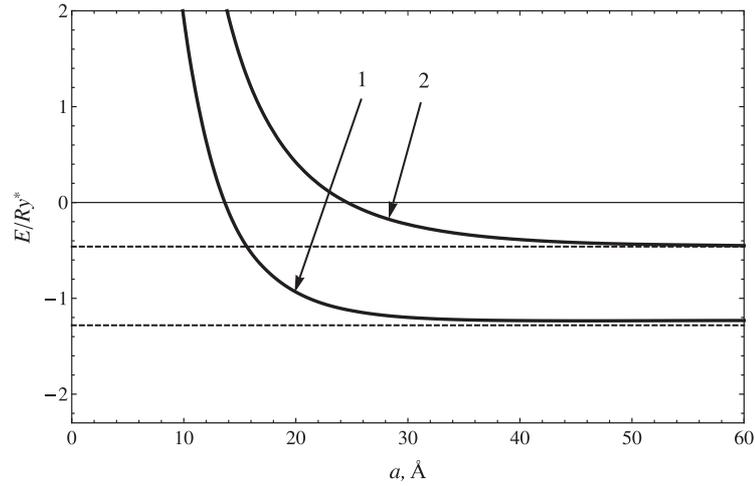}
}
\caption{The energy of the central acceptor impurity in the  QD.} \label{fig1}
\end{figure}

Specific calculations were made for CdSe QD. We take the following basic parameters of the crystal that
form the QD: $E_\mathrm{g}=1.841$~eV, $\Delta=0.420$~eV~\cite{Madelung}, $\gamma_1=1.66$, $\gamma=0.41$~\cite{Laheld}, $\varepsilon=9.53$~\cite{Mend}. Initially, the energy of the central acceptor impurity $D=0$ was determined using the Ritz variational method. Calculation results of the energy as the function of the QD radius are presented in figure~\ref{fig1}.

In figure~\ref{fig1}, there are two curves which correspond to the ground state (first type of states $f=3/2$)~-- curve~1, and to the first excited state (second type of states $f=3/2$)~-- curve~2. Horizontal dashed lines denote the energies of the acceptor impurity in the bulk crystal CdSe~\cite{Bald}. Those energies saturate at small QD radii ($a<40$~\AA). This is due to a small effective Bohr radius ($a^*_\mathrm{b}=8.38$~\AA). As expected, there remains a degeneracy by the quantum number $M$.

\begin{figure}[!h]
\centerline{
\includegraphics[width=0.65\textwidth]{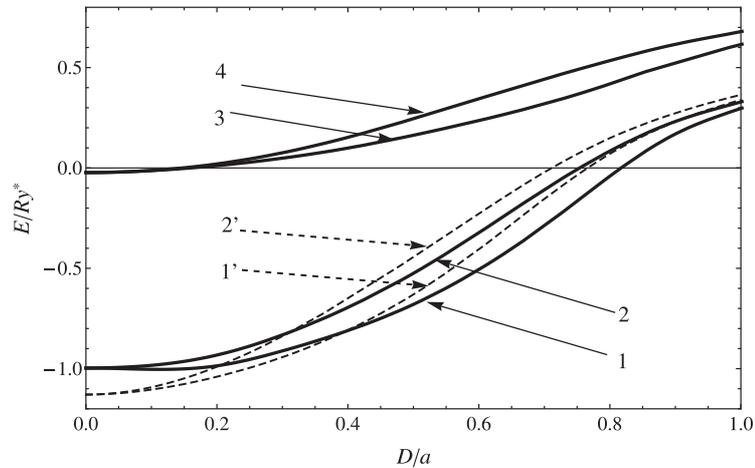}
}
\caption{The energy of the off-central acceptor impurity in the QD as a function of the impurity location. The QD radius is $a=25$~\AA.} \label{fig2}
\end{figure}
The calculations of the energy of the off-central acceptor impurity were made using the Ritz variational method (only ground state) and by the linear expansion over the wave function without the impurity (many states). For a detailed analysis, further calculations $E=E(D)$ were made for the intermediate QD radius ($a=25$~\AA). Calculation results of the energy of the off-central acceptor impurity are presented in  figure~\ref{fig2}. Solid curves denote the energy of the acceptor impurity which was calculated using the method of linear expansion over the wave function without impurity, dashed curves denote the same energies that were calculated by the Ritz variational method.
As mentioned above, the variational functional depends on the quantum number $M$: $F = F\left( {{Q_{\mathrm{I}}},{W_{\mathrm{I}}},{\alpha _{\mathrm{I}}},M} \right)$. Therefore, energy of the off-central impurity also formally depends on $M$. But specific calculations showed that the energy of the off-central impurity depends only on $|M|$. Thereby, if the ion of the impurity shifts from the QD center, the energy level of the ground state splits into two levels: with $|M|=3/2$ (curve~$1'$) and with $|M|=1/2$ (curve~$2\,'$). The cause of this dependence on $|M|$ can be explained by the violation of the spherical symmetry of the problem, and by the preservation of the cylindrical symmetry. A similar splitting of energy levels was obtained for the donor impurity~\cite{Boic4}. However, due to the fact that the ground state of the donor impurity does not degenerate in the single-band model, there is no splitting of the ground state energy of the donor impurity in contrast to the acceptor impurity.

Apart from the Ritz variational method, there was used a method of linear expansion over the wave function without impurity. The first 74 terms were used in calculations. For small $D/a$, this method with a specified number of terms gives the energy which differs  by 10 percent from the Ritz method. For a large $D/a$ everything changes to the contrary. Since for a small $D/a$, the symmetry of the problem is close to spherically symmetric problem, the contribution of $W_{ji}$ to degenerated states becomes small and in the case of $D/a=0$ it is equal to zero. In this regard, for small $D/a$, the number of terms in the expansion should be increased four times which dramatically increases the computation time. Therefore, for small displacements of the impurity, it is better to use the proposed Ritz variational method, while for a large $D/a$, the method of decomposition turns out to be better for calculations. As an example, for $D/a$ we got the following results:
\begin{eqnarray}
    {\Psi _{1, - 3/2}} &=& 0.609116\psi _{3/2, - 3/2;1,0}^{\mathrm{I}} - 0.644407\psi _{3/2, - 3/2;1;1}^{\mathrm{II}} - 0.261659\psi _{3/2, - 3/2;2,0}^{\mathrm{I}}  \nonumber \\
    &&{}+ 0.163952\psi _{5/2, - 3/2;1,1}^{\mathrm{I}} - 0.287506\psi _{5/2, - 3/2;1,2}^{\mathrm{II}} - 0.15593\psi _{5/2, - 3/2;2,1}^{\mathrm{I}} + \ldots , \qquad \\[2ex]
%
    {\Psi _{1,3/2}} &=& 0.609116\psi _{3/2,3/2;1,0}^{\mathrm{I}} + 0.644407\psi _{3/2,3/2;1;1}^{\mathrm{II}} - 0.261659\psi _{3/2,3/2;2,0}^{\mathrm{I}}  \nonumber \\
      &&{}+ 0.163952\psi _{5/2,3/2;1,1}^{\mathrm{I}} + 0.287506\psi _{5/2,3/2;1,2}^{\mathrm{II}} - 0.15593\psi _{5/2,3/2;2,1}^{\mathrm{I}} + \ldots.
\end{eqnarray}
The following terms are an order of magnitude smaller. Those two states have the same energy ${E_{1,\left| {3/2} \right|}} =  - 0.50588$~Ry$^*$. Similar energies for other values of $D/a$ are indicated by a curve~1 (figure~\ref{fig2}). States $\Psi _{1, - 1/2}$ and $\Psi _{1,1/2}$ have the energy $E_{1,|3/2|}=- 0.32289$~Ry$^*$. Thus, the curve~1 (figure~\ref{fig2}) denotes the energy $E_{1,|3/2|}$ of degenerated states $\Psi_{1,-3/2}$, $\Psi_{1,3/2}$; the curve~2 (figure~\ref{fig2}) denotes $E_{1,|1/2|}$ of $\Psi_{1,-1/2}$, $\Psi_{1,1/2}$; the curve~3 (figure~\ref{fig2}) denotes $E_{2,|3/2|}$ of $\Psi_{2,-3/2}$, $\Psi_{2,3/2}$; the curve~4 (figure~\ref{fig2}) denotes $E_{2,|1/2|}$ of $\Psi_{2,-1/2}$, $\Psi_{2,1/2}$.

In practice, it is very difficult to get a single isolated QD. In most cases, the set of QD are received which can be characterized by the average QD radius and by dispersion. If the average QD radius is 25~\AA \ and the dispersion is 16\%, the splitting by the quantum number $|M|$  vanishes. However, the energy of states 1 and 2 does not intersect each other. This conclusion is important assuming that the dipole transitions between states with the energies $E_{1,|3/2|}$ and $E_{1,|1/2|}$ are forbidden, between $E_{1,|3/2|}$, $E_{2,|3/2|}$ are permitted, and between $E_{1,|1/2|}$, $E_{2,|1/2|}$ are permitted too. These optical transitions are calculated in the section of the paper that follows.

\section{The interlevel transition of the hole of the acceptor impurity}

We have considered the case of the QD irradiated by a linearly polarized light along $z$ direction. The dipole transition matrix element between two states is calculated using the function~(\ref{WaveFuctionExpansion}). However, coefficients $c_i$ are the wave function in the $\psi _i^0$ representation, and we wrote the dipole transition matrix element in the same representation
\begin{equation}\label{dipoleRpres}
    {d^0}_{ij} = \re{z^0}_{ij} = \re\left\langle {\psi _i^0\left| z \right|\psi _j^0} \right\rangle.
\end{equation}
\begin{figure}[!h]
\centerline{
\includegraphics[width=0.65\textwidth]{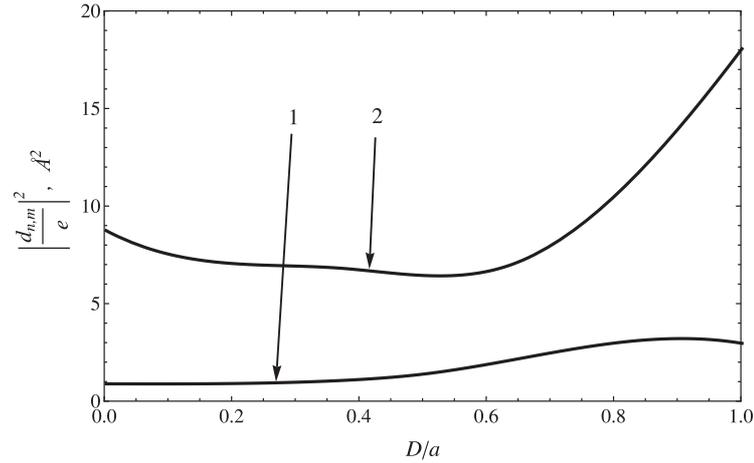}
}
\caption{Squared matrix element of the dipole moment of the interlevel transitions of the off-central acceptor impurity as a function of the location of the impurity in the QD. The QD radius is $a=25$~\AA.} \label{fig3}
\end{figure}
\begin{figure}[!h]
\centerline{
\includegraphics[width=0.65\textwidth]{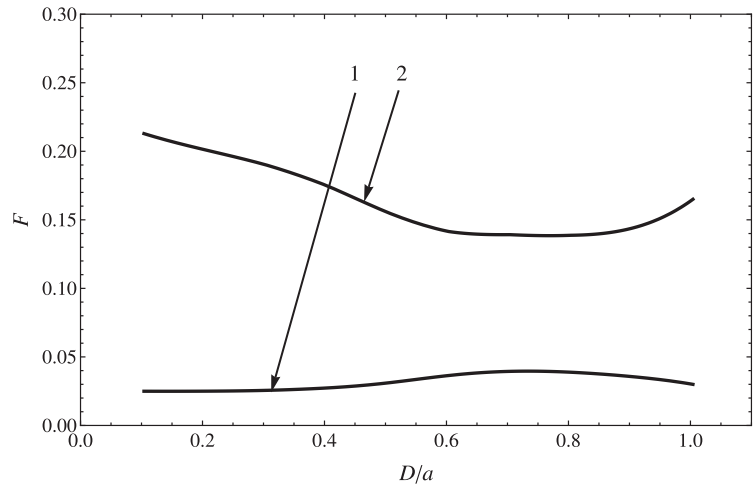}
}
\caption{The oscillator strength of interlevel transitions. 1~-- ${F_{1,1/2 \to 2,1/2}} = {F_{1, - 1/2 \to 2, - 1/2}}$\,, 2~-- ${F_{1,3/2 \to 2,3/2}} = {F_{1, - 3/2 \to 2, - 3/2}}$\,. } \label{fig4}
\end{figure}
After determining the matrix elements of the matrix $d^0$, the dipole transition matrix element of the off-central impurity was defined
\begin{equation}\label{dipole}
    {d_{nm}} = {\left( {{C_n}} \right)^T}{d^{\,0}}{C_m}\,,
\end{equation}
where $C_n$ is the vector which consists of coefficients $c_i$ for the $n$-state of the acceptor impurity. The calculation of the dipole momentum shows that transitions are possible when the quantum number $M$ does not change. That is why ${d_{1, - 3/2;2, - 3/2}} \ne 0$ (${\Psi _{1, - 3/2}} \leftrightarrow {\Psi _{2, - 3/2}}$), ${d_{1,3/2;2,3/2}} \ne 0$ ($\Psi _{1,3/2} \leftrightarrow {\Psi _{2,3/2}}$), ${d_{1, - 1/2;2, - 1/2}} \ne 0$ (${\Psi _{1, - 1/2}} \leftrightarrow {\Psi _{2, - 1/2}}$), ${d_{1,1/2;2,1/2}} \ne 0$ (${\Psi _{1,1/2}} \leftrightarrow {\Psi _{2,1/2}}$); ${\left| {{d_{1, - 3/2;2, - 3/2}}} \right|^2} = {\left| {{d_{1,3/2;2,3/2}}} \right|^2}$, \linebreak ${\left| {{d_{1, - 1/2;2, - 1/2}}} \right|^2} = {\left| {{d_{1,1/2;2,1/2}}} \right|^2}$. Graphics of the matrix element of the dipole moment of the interlevel transitions are shown in figure~\ref{fig3}.
The curve~1 represents ${\left| {{d_{1, - 1/2;2, - 1/2}}} \right|^2} = {\left| {{d_{1,1/2;2,1/2}}} \right|^2}$, the curve~2 represents ${\left| {{d_{1, - 3/2;2, - 3/2}}} \right|^2} = {\left| {{d_{1,3/2;2,3/2}}} \right|^2}$. In addition, the oscillator strength of interlevel transitions was obtained (figure~\ref{fig4}) using the formula
\begin{equation}\label{oscilator}
    {F_{1,M \to 2,M'}} = \frac{{2{m_0}}}{{{\re^2}{\hbar ^2}{\gamma _1}}}\left( {{E_{2,\left| {M'} \right|}} - {E_{1,\left| M \right|}}} \right){\left| {{d_{1,M;2,M'}}} \right|^2}.
\end{equation}
The linear optical absorption coefficient which is caused by the interlevel optical transition was defined based on the expression~\cite{Vahdani,Rezaei2}:
\begin{equation}\label{absorption}
    {\alpha _{1,\left| M \right|;2,\left| M \right|}}\left( \omega  \right) = \omega \sqrt {\frac{{{\mu _0}}}{{{\varepsilon _0}\varepsilon }}} \frac{{\sigma \left( {{{\left| {{d_{1,\left| M \right|;2,\left| M \right|}}} \right|}^2} + {{\left| {{d_{1, - \left| M \right|;2, - \left| M \right|}}} \right|}^2}} \right)\hbar \Gamma }}{{{{({E_{2,\left| M \right|}} - {E_{1,\left| M \right|}} - \hbar \omega )}^2} + {{\left( {\hbar \Gamma } \right)}^2}}}\,,
\end{equation}
where $\omega$  is the frequency of the light, $\mu_0$ is magnetic constant, $\hbar \Gamma$ is the relaxation rate caused by the electron-phonon interaction and some other factors of the scattering, $|M|=1/2, 3/2$. The electron density in the QD $\sigma$ is chosen on the assumption that the QD has one hole (impurity hole). That is why $\sigma  = 3/\left( {4\pi {a^3}} \right)$. We assume that the QD is under low temperature, and its surface is ideally spherical. That is why $\hbar \Gamma$ can be estimated as the energy level width which is caused by the scattering on acoustic phonons. If the temperature of the system is $T\approx 20$~K, then $\hbar \Gamma  = k_\mathrm{B} T \approx 1.7$~meV, where $k_\mathrm{B}$ is Boltzman constant.

\begin{figure}[!h]
\centerline{
\includegraphics[width=0.65\textwidth]{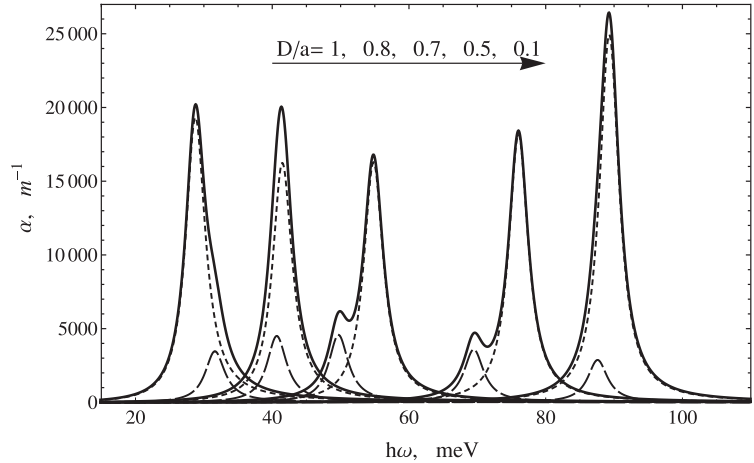}
}
\caption{The optical absorption coefficient. Dashed curves represent ${\alpha _{1,1/2;2,1/2}}$\,, dotted curves represent ${\alpha _{1,3/2;2,3/2}}$\,. The total absorption coefficient is denoted by solid curves.} \label{fig5}
\end{figure}
The dependence of the optical absorption coefficient on the impurity ion position is presented in figure~\ref{fig5}. The curves of light absorption are of the form of Lorenz curve. That is why the squares under those curves are proportional to the oscillator strength of interlevel transitions. However ${F_{1,3/2 \to 2,3/2}} > {F_{1,1/2 \to 2,1/2}}$, then ${\alpha _{1,3/2;2,3/2}} > {\alpha _{1,1/2;2,1/2}}$ for all frequencies. It is seen from figure~\ref{fig5} that the displacement of the impurity from QD center causes the shift of absorption bands into a lower energy region. This shift is caused by a decrease of the distance between energy levels. Similar results were obtained for a donor impurity~\cite{Boic4}. However, the difference lies in the absence of the ground state of the donor impurity for degeneration as compared with the acceptor impurity. In addition, it should be noted that the absorption bands that correspond to a transition between the states with energies $E_{1,|3/2|}\rightarrow E_{2,|3/2|}$ for the intermediate $D/a$ are shifted into a high energy region as compared with transitions $E_{1,|1/2|}\rightarrow E_{2,|1/2|}$. The difference between those transition energies explain the fact that in figure~\ref{fig5}, for $D/a=0.5; 0.7$, there is seen a ``structure'' in $\alpha=\alpha(\omega)$ dependence. For a large or small $D/a$, the function $\alpha=\alpha(\omega)$ is of a Lorenz form. In real heterostructures, there is observed a dispersion by the QD size. Calculations show that if the dispersion is more than 15\%, the half-width of absorption bands will be larger. That is why the structure cannot be observed.

\newpage
\section{Summary}

In the proposed paper, there was made a theoretical study of the energy spectrum of the off-central acceptor impurity based on the Luttinger model. It is possible to determine:
\begin{itemize}
  \item the dependence of the energy on the acceptor impurity position in the QD;
  \item a partial removal of the degeneracy of energy levels by a quantum number $|M|$;
  \item the QD optical absorption coefficient with the off-central impurity shows that the displacement of the impurity from the QD center causes the shift of absorption bands into the low energy range. It is also proved that some obtained results are qualitatively similar to the results of the off-central donor impurity.
\end{itemize}

\ukrainianpart

\title{Нецентральна акцепторна домішка у сферичній квантовій точці}

\author{В.І.~Бойчук, Р.Я.~Лешко, І.В.~Білинський, Л.М.~Турянська}

\address{Кафедра теоретичної фізики, Дрогобицький державний педагогічний університет ім.~Івана Франка, \\ вул. Стрийська, 3, 82100, Дрогобич, Львівська обл., Україна
}

\makeukrtitle

\begin{abstract}
\tolerance=3000%
У рамках сферичного наближення багатозонної моделі Латтінджера проведено обчислення енергетичного спектру дірки за наявності іона акцепторної домішки у сферичній квантовій точці. Досліджено за\-леж\-ність енергетичних рівнів дірки від розташування домішки у наносистемі. Проаналізовано вплив положення домішки на дипольний момент та силу осцилятора міжрівневих переходів. Визначено коефіцієнт поглинання світла, зумовлений міжрівневими переходами дірки.
\keywords коефіцієнт поглинання, нецентральна акцепторна домішка

\end{abstract}

\end{document}